\documentclass{article}
\usepackage{arxiv}
\usepackage[utf8]{inputenc} % allow utf-8 input
\usepackage[T1]{fontenc}    % use 8-bit T1 fonts
\usepackage{hyperref}       % hyperlinks
\usepackage{url}            % simple URL typesetting
\usepackage{booktabs}       % professional-quality tables
\usepackage{amsfonts}       % blackboard math symbols
\usepackage{nicefrac}       % compact symbols for 1/2, etc.
\usepackage{microtype}      % microtypography
\usepackage{lipsum}		% Can be removed after putting your text content
\usepackage{graphicx}
\usepackage{threeparttable}
\usepackage{doi}
\usepackage{multirow}
\usepackage{multicol}
\usepackage{xcolor}
\usepackage{subcaption}
\usepackage[normalem]{ulem}
\usepackage{amsmath,amsfonts,bm}
\usepackage{amssymb,amsthm}
\usepackage{enumitem}
\usepackage{adjustbox}

\usepackage[
    backend=biber,
    style=numeric,
]{biblatex}
\newcommand{\citep}[1]{\parencite{#1}}
\setlist[itemize,1]{leftmargin=\dimexpr 18pt}
\setlist[enumerate,1]{leftmargin=\dimexpr 18pt}

\title{
\raisebox{-0.1\height}{\includegraphics[width=0.036\textwidth]{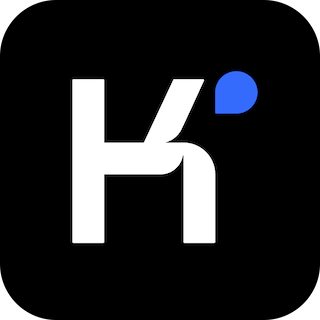}} %
Kimi-Audio Technical Report}
\author{\large Kimi Team \\}

\date{}

\renewcommand{\headeright}{Technical Report}
\renewcommand{\undertitle}{}
\renewcommand{\shorttitle}{\raisebox{-0.12\height}{\includegraphics[width=0.02\textwidth]{figures/logo.png}}
Kimi-Audio}

\begin{document}
\large

\maketitle

\vspace{-0.5cm}

\begin{abstract}
We present Kimi-Audio, an open-source audio foundation model that excels in audio understanding, generation, and conversation. We detail the practices in building Kimi-Audio, including model architecture, data curation, training recipe, inference deployment, and evaluation. Specifically, we leverage a $12.5$Hz audio tokenizer, design a novel LLM-based architecture with continuous features as input and discrete tokens as output, and develop a chunk-wise streaming detokenizer based on flow matching.  
We curate a pre-training dataset that consists of more than $13$ million hours of audio data covering a wide range of modalities including speech, sound, and music, and build a pipeline to construct high-quality and diverse post-training data. Initialized from a pre-trained LLM, Kimi-Audio is continual pre-trained on both audio and text data with several carefully designed tasks, and then fine-tuned to support a diverse of audio-related tasks. Extensive evaluation shows that Kimi-Audio achieves state-of-the-art performance on a range of audio benchmarks including speech recognition, audio understanding, audio question answering, and speech conversation. We release the codes, model checkpoints, as well as the evaluation toolkits in \url{https://github.com/MoonshotAI/Kimi-Audio}. 

\begin{figure}[htb!]
\centering
\hspace{1.3cm}
\includegraphics[width=0.7\textwidth]{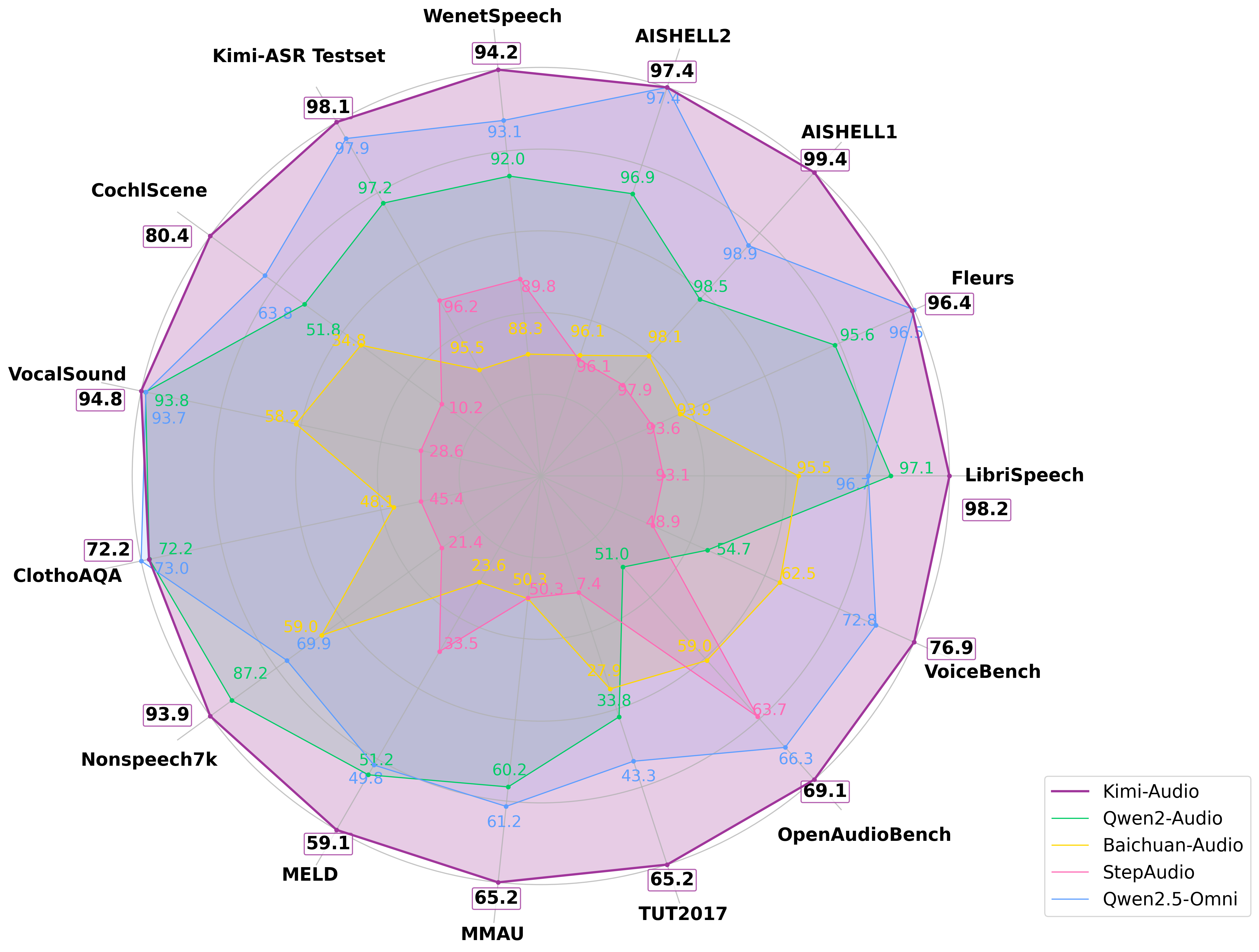}
\caption{Performance of Kimi-Audio and previous audio langauge models including Qwen2-Audio~\citep{chu2024qwen2}, Baichuan-Audio~\citep{li2025baichuan}, Step-Audio~\citep{huang2025step}, and Qwen2.5-Omni~\citep{xu2025qwen2} on various benchmarks.}
\end{figure}

\end{abstract}

\section{Introduction}
Audio plays an indispensable role in human daily life, such as environment perception, speech conversation, emotion expression, and music appreciation, and is an important topic in artificial general intelligence. Traditional audio modeling, constrained by the development of artificial intelligence, handles each audio processing task (e.g., speech recognition, emotion recognition, sound event detection, and speech conversation) separately. However, audio is naturally sequential and speech has strict correspondence with text, which makes it suitable to take advantage of the rapid progress in large language models (LLMs) in audio modeling. Just as natural language processing has experienced, audio processing evolves quickly from separate models for separate tasks to a universal model handling a variety of tasks. 

For example, pioneer works introduce language models in audio generation~\citep{borsos2023audiolm,yang2023uniaudio}, audio understanding~\citep{chu2023qwen,tang2023salmonn,cheng2024}, speech recognition~\citep{radford2023robust,zhang2023google}, speech synthesis~\citep{wang2023neural,ye2025llasa}, and end-to-end speech conversation~\citep{hurst2024gpt,defossez2024moshi}. However, previous works fall short of building a universal audio foundation model for a variety of audio processing tasks in several aspects: 1) not universal but only focus on a specific type of tasks, such as audio understanding~\citep{chu2023qwen,chu2024qwen2,geng2025osum,tang2023salmonn,zhao24h,kong2024audio}, audio generation~\citep{liu2024convincing,yang2023uniaudio}, or speech conversation~\citep{defossez2024moshi,zeng2024glm}; 2) not much emphasis on audio pre-training but only fine-tuning an LLM on downstream audio tasks~\citep{chu2023qwen,tang2023salmonn,cheng2024}; 3) no access to source codes and checkpoints, with limited value to the community~\citep{hurst2024gpt,chen2025minmo}.

In this report, we present Kimi-Audio, an open-source audio foundation model that handles a variety of audio processing tasks. We detail our effort in building a state-of-the-art (SOTA) audio foundation model in three espects: architecture, data, and training. 
\begin{itemize}
\item Architecture. Our model consists of three components: an audio tokenizer and detokenizer as audio I/O, and an audio LLM as the core processing part (see Section~\ref{sec:arc_overview}). We use discrete semantic audio tokens as the basic representation for both the input and output of the audio LLM. Meanwhile, we concatenate the semantic audio token with continuous acoustic vectors in the input to enhance perception capability, and concatenate with discrete text tokens in the output to enhance the generation capability. In this way, we can achieve good audio perception and generation capabilities at the same time, facilitate universal audio modeling. We reduce the number of token per second in audio to bridge the gap between text and audio sequence and set the compression rate of both the semantic and acoustic audio tokens as $12.5$Hz. The detailed design of the audio tokenizer for both discrete semantic tokens and continuous acoustic vectors, as well as the generation of both discrete semantic tokens and text tokens are introduced in Section~\ref{sec:audio_tokenizer} and ~\ref{sec:audio_llm} respectively.
\item Data. To achieve SOTA universal audio modeling, we need to pre-train the model on a large amount of audio data to see diverse scenarios. To this end, we crawl and process a large-scale audio pre-training dataset. We develop a data processing pipeline consisting of speech enhancement, diarization, transcription, filtering, etc, to enable high data quality (see Section~\ref{sec:data/pretraining}). To support diverse audio processing tasks, we curate a large amount of task-specific data for supervised fine-tuning (SFT). We demonstrate an economic way to construct most of SFT data with pure open and accessible data sources and processing tools to achieve SOTA performance, without relying on any data purchase (see Section~\ref{sec:data/sftdata}). 
\item Training. To achieve good audio understanding/generation capability while maintaining high knowledge capacity and intelligence, we initialize the audio LLM with a pre-trained LLM, and carefully design a series of pre-training tasks to fully learn the audio data and bridge the gap between text and audio. Specifically, the pre-training tasks can be divided into three categories: 1) text-only and audio-only pre-training, which aims to learn the knowledge from text and audio domains separately; 2) audio-to-text mapping, which encourages the conversion between audio and text; 3) audio-text interleaving, which further bridge the gap between text and audio (see Section~\ref{sec:training/pretraining}). In the supervised fine-tuning stage, we develop a training recipe to improve fine-tuning efficiency and task generalization (see Section~\ref{sec:training/sft}).  
\end{itemize}

Furthermore, we introduce the practices for deploying and serving our audio foundation model for inference in Kimi APP, as described in Section~\ref{sec:deploy}. Evaluating and benchmarking an audio foundation model in various downstream tasks such as speech recognition, audio understanding, and speech conversation is challenging. We encounter tricky issues in fairly comparing different audio models, such as non-standardized metric, evaluation protocol, and inference hyper-parameters. Therefore, we develop an evaluation toolkit that can faithfully evaluate audio LLMs on comprehensive benchmarks (see Section~\ref{sec:eval_tool}). We open-source this toolkit to facilitate a fair comparison in the community. 

Based on our evaluation toolkit, we conduct a comprehensive evaluation of Kimi-Audio and other audio LLMs on a variety of audio benchmarks (see Section~\ref{sec:eval_result}). Evaluation results demonstrate that Kimi-Audio achieves SOTA performance in a series of audio tasks, including speech recognition, audio understanding, audio-to-text chat, and speech conversation. We open source the codes and checkpoints of Kimi-Audio, as well as the evaluation toolkit in \url{https://github.com/MoonshotAI/Kimi-Audio}, to boost the development of the community.

\section{Architecture}
\label{sec:arc}
\begin{figure}[t]
\centering
\includegraphics[width=0.8\textwidth]{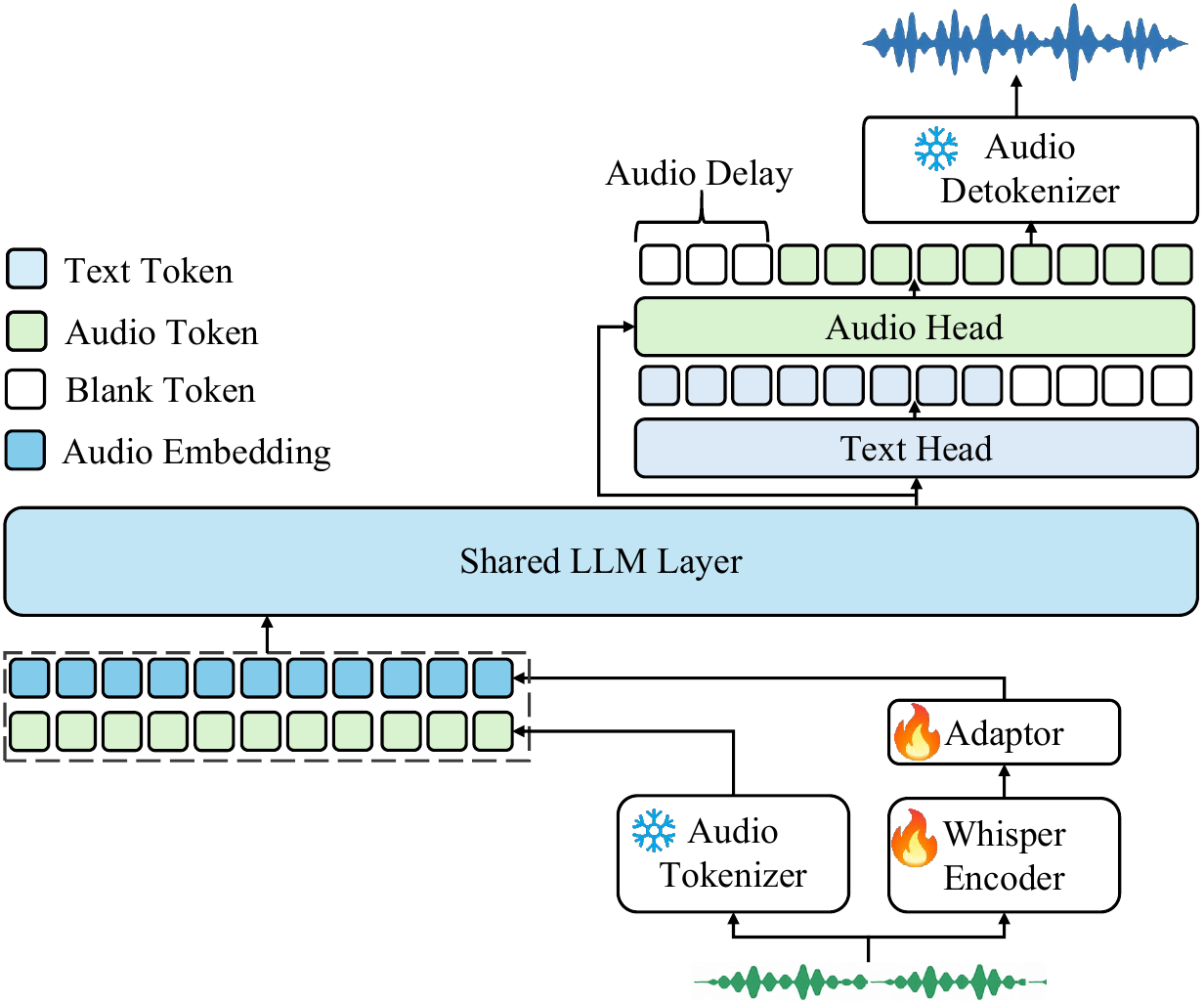} 
\caption{Overview of the Kimi-Audio model architecture: (1) an audio tokenizer that extracts discrete semantic tokens and a Whisper encoder that generates continuous acoustic features; (2) an audio LLM that processes audio inputs and generates text and/or audio outputs; (3) an audio detokenizer converts audio tokens into waveforms.}
\label{fig:overall_pipeline}
\end{figure}
\subsection{Overview}
\label{sec:arc_overview}
Kimi-Audio is an audio foundation model designed to perform comprehensive audio understanding, generation, and conversation tasks within a unified architecture. As illustrated in Figure~\ref{fig:overall_pipeline}, our system comprises three primary components: (1) an audio tokenizer that converts input audio into discrete semantic tokens derived through vector quantization with a $12.5$Hz frame rate. The audio tokenizer additionally extracts continuous acoustic vectors to enhance perception capability. (2) an audio LLM that generates semantic tokens together with text token to improve the generation capability, featuring shared transformer layers that process multimodal inputs before branching into specialized parallel heads for text and audio generation; and (3) an audio detokenizer that converts the discrete semantic tokens predicted by the audio LLM back into coherent audio waveforms using a flow matching approach. This integrated architecture enables Kimi-Audio to seamlessly handle diverse audio-language tasks from speech recognition and understanding to speech conversation within a single unified model framework.

\subsection{Audio Tokenizer}
\label{sec:audio_tokenizer}
Our audio foundation model employs a hybrid audio tokenization strategy, integrating discrete semantic tokens and complementary continuous vectors of acoustic information to effectively represent speech signals for downstream tasks. This tokenization allows the model to leverage the efficiency and semantic focus of discrete tokens while benefiting from the rich acoustic details captured by continuous representations.

We incorporate the discrete semantic tokens proposed by GLM-4-Voice~\citep{zeng2024glm}. This component utilizes a supervised speech tokenizer derived from an automatic speech recognition (ASR) model. By introducing a vector quantization layer within the whisper encoder architecture~\citep{radford2023robust}, we can transform continuous speech representations into a sequence of discrete tokens at a low frame rate (i.e. $12.5$Hz) using a single codebook.

Complementing the discrete semantic tokens, we incorporate a continuous feature representation derived from a pre-trained whisper model~\citep{radford2023robust} to enhance the perception capability of our model. Since the whisper feature has a frame rate of $50$Hz, we additionally introduce an adaptor upon the whisper feature extractor to downsample the feature from $50$Hz to $12.5$Hz. The downsampled features are added to the embeddings of discrete semantic tokens to serve as the input of the audio LLM.

By combining discrete semantic tokens with continuous whisper features, our model benefits from efficient, semantically grounded representation and detailed acoustic modeling, providing a comprehensive foundation for diverse audio processing tasks.

\subsection{Audio LLM}
\label{sec:audio_llm}
The core of our system is an audio LLM designed to process the audio representations generated by the tokenization strategy described in Section~\ref{sec:audio_tokenizer} and produce multimodal outputs, which include the discrete semantic tokens of audio and the corresponding text tokens to improve the generation capability.

To enable the model to generate both audio semantic tokens and the corresponding textual responses, we adapt the standard LLM architecture by structuring it into components with shared and specialized functions. A significant portion of the original transformer bottom layers, i.e., the first several layers, are utilized as shared layers. These layers process the input sequence and learn cross-modal representations, integrating information from both text and audio modalities present in the input or context. Based on these shared layers, the architecture diverges into two parallel heads containing transformer layers. The first head is a text head that is specifically responsible for autoregressively predicting text tokens, forming the textual output of the model. The second head is an audio head to predict the discrete audio semantic tokens. These predicted audio tokens are subsequently passed to an audio detokenizer module to synthesize the final output audio waveform.

To take advantage of the strong language capabilities of the pre-trained text LLMs~\citep{qwen2.5,grattafiori2024llama,deepseekai2024deepseekv3technicalreport}, the parameters of the shared transformer layers and the text head are initialized directly from the weights of the pre-trained text LLM. The audio head layers are initialized randomly. This initialization strategy ensures that the model retains robust text understanding and generation capabilities while learning to effectively process and generate audio information.

\subsection{Audio Detokenizer}
\label{sec:audio_detokenizer}
The audio detokenizer aims to generate high-quality and expressive speech conditioned on discrete semantic audio tokens. We employ the same detokenizer architecture as in MoonCast~\citep{ju2025mooncast}, which contains two parts: 1) a flow-matching module which converts $12.5$Hz semantic tokens to $50$Hz mel-spectrograms; 2) a vocoder which generates waveforms from mel-spectrograms. To reduce speech generation latency, we design a chunk-wise streaming detokenizer. Intuitively, we can split the semantic tokens into chunks and decode them separately, which, however, in our preliminary experiments, faces an intermittent issue in the chunk boundaries. Thus, we propose a chunk-wise autoregressive streaming framework with a look-ahead mechanism.

\textbf{Chunk-wise Autoregressive Streaming Framework.}
We split the audio into chunks (e.g., $1$ second per chunk): $\{c_1, c_2, ..., c_i, ..., c_N\}$, where $N$ is the number of chunks. Firstly, to match the sequence length between semantic tokens ($12.5$Hz) and mel-spectrograms ($50$Hz), we upsample the semantic tokens by $4$x rate. Secondly, we apply a chunk-wise causal mask during training and inference, i.e., for chunk $c_i$, all previous chunks $c_j$ with $j<i$ are prompts. We denote chunk $c_i$'s mel-spectrograms as $m_i$ and the corresponding discrete semantic audio tokens as $a^d_i$. The flow-matching model's forward step will mix $m_i$ with Gaussian noise and the backward step will remove noise to obtain clean $m_i$ with condition $a^d_i$ and prompt $c_j$, where $j<i$, and $c_j$ contains both $m_j$ and $a^d_j$. With this design, during inference, when the LLM generates a chunk, we employ the flow-matching model to detokenize it to obtain the mel-spectrograms. Finally, we apply a BigVGAN~\citep{lee2022bigvgan} vocoder to generate wavforms for each chunk.

\textbf{Look-Ahead Mechanism.}
With a preliminary study, we find that the generated audio in the boundaries of chunks still has an intermittent issue. Although a long range of history context has been seen during the diffusion denoising process, the future context of the boundary position cannot be seen due to the nature of block-wise causal attention, which causes the degradation of quality. Thus, we propose a look-ahead mechanism. In detail, for chunk $c_i$, we take the future $n$ (e.g. $4$) semantic tokens from chunk $c_{i+1}$ and concatenate them to the end of $c_i$ to form $\hat{c}_i$. Then we detokenize $\hat{c}_i$ to generate the mel-spectrograms, but only retain the mel-spectrograms corresponding to $c_i$. This mechanism is training-free and will only delay the generation of the first chunk by $n$ tokens.

\section{Data}

\subsection{Pre-Training Data}
\label{sec:data/pretraining}

\begin{figure}
    \centering
    \includegraphics[width=1.0\columnwidth]{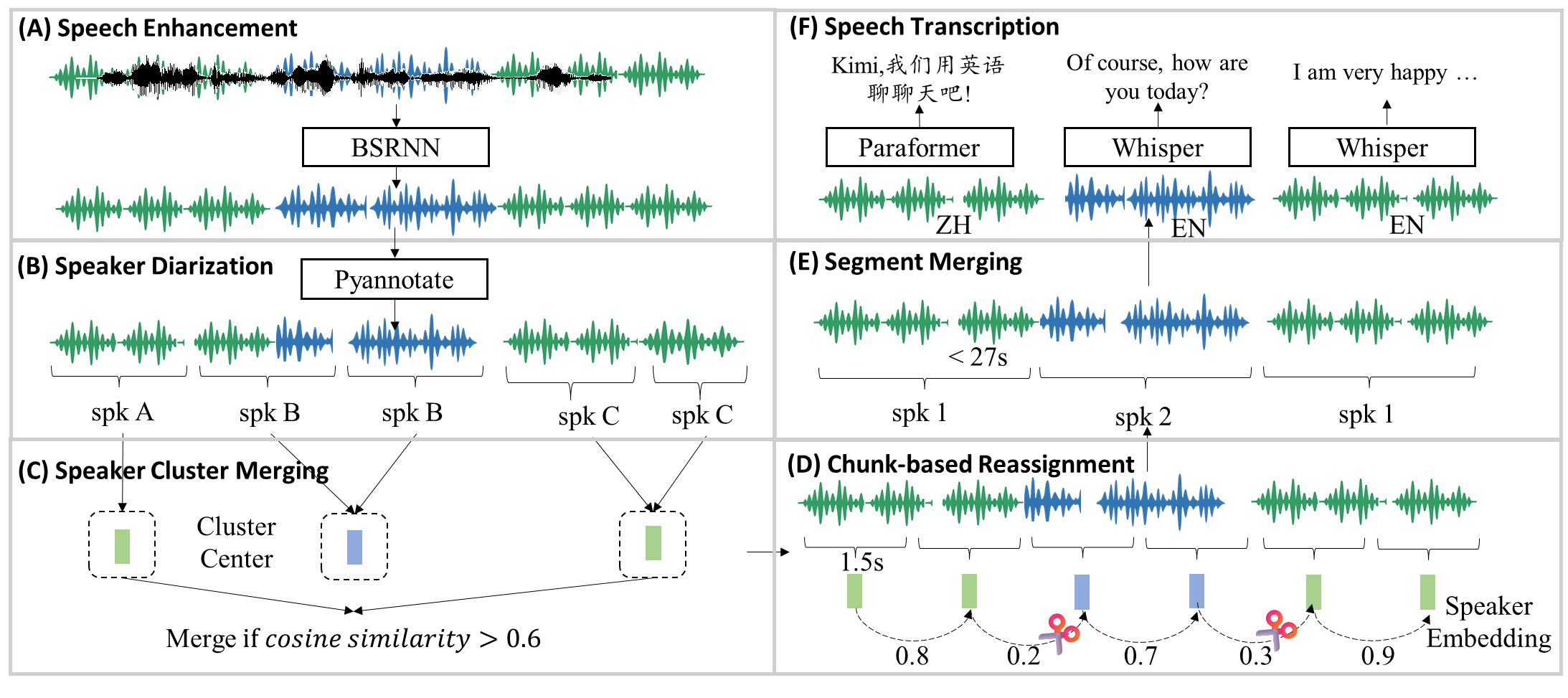}
    \caption{Processing pipeline for the audio pre-training data.}
    \label{fig:pretrain_data_pipeline}
\end{figure}
Our pre-training corpus comprises both unimodal (text-only, audio-only) and multimodal (text-audio) data. 
The audio-only pre-training data covers a wide range of real-world scenarios, including audiobooks, podcasts, and interviews, and consists of approximately $13$ million hours of raw audio containing rich acoustic events, music, environmental sound, human vocalization, and multilingual information. The details of the text-only pre-training data can be found in~\citep{kimiteam2025kimik15scalingreinforcement}.

Most audio corpus contains only raw audio without corresponding transcriptions, language types, speaker annotations, and segmentation boundaries. In addition, the raw audio often contains undesired artifacts such as background noise, reverberation, and speaker overlap. 

Inspired by previous work~\citep{ju2025mooncast,yu2024autoprep,he2024emilia}, we develop an efficient automatic audio data processing pipeline to generate high-quality annotations, resulting in our multimodal (audio-text) data.

Compared to previous data processing pipelines that primarily focus on generating high-quality short audio segments without contextual information, our pipeline is designed to provide long-form audio annotations with consistent long-range context.
The pipeline includes the following key components in a step-by-step manner, as shown in Figure~\ref{fig:pretrain_data_pipeline} and described as follows.

\textbf{Speech Enhancement.} To suppress undesired background noise and reverberation, we develop a speech enhancement model based on the Band-Split RNN (BSRNN) architecture~\citep{BSRNNLY}, as shown in Figure~\ref{fig:pretrain_data_pipeline}(A). Following the same hyper-parameter configuration as in~\citep{yu2022high}, the model is applied to perform $48$kHz speech enhancement.
Empirically, we find that speech enhancement will remove the environmental sound and music, which can be harmful to audio understanding. Thus, we randomly choose original or enhanced audio with a ratio of $1:1$ in the pre-training stage.

\textbf{Segmentation by Diarization.} We employ a diarization-driven approach to segment long-form audio. We utilize the PyAnnote toolkit\footnote{\url{https://github.com/pyannote/pyannote-audio}} for speaker diarization (Figure~\ref{fig:pretrain_data_pipeline}(B)), which segments the audio and assigns speaker labels. However, the raw output is sub-optimal, and thus we develop a post-processing pipeline to address the issues in the previous segmentation results:
\begin{itemize}
\item \textbf{Speaker Cluster Merging.} We observe that PyAnnote sometimes assigns multiple speaker labels to the same actual speaker, which results in speaker fragmentation. We compute representative speaker embeddings for each initial cluster and merge pairs of clusters whose embeddings have a cosine similarity greater than $0.6$, as shown in Figure~\ref{fig:pretrain_data_pipeline}(C).

\item \textbf{Chunk-based Reassignment.} The initial diarization occasionally produced segments containing multiple speakers. To purify the segments, 1) we first divide all segments into $1.5$-second chunks, and then 2) for each pair of adjacent chunks, if their cosine similarity is below $0.5$, we treat them as belonging to different speakers and reassign each chunk to the speaker cluster with the highest similarity, as shown in Figure~\ref{fig:pretrain_data_pipeline}(D).

\item \textbf{Segment Merging.} The initial diarization could result in segments of highly variable and sometimes impractical lengths (shorter than $1$s or longer than $100$s). So we iteratively merge adjacent segments labeled with the same speaker (after the reassignment step).
The merging process terminates if the accumulated segment length exceeds $27$ seconds or the silence gap between two segments is greater than $2$ seconds, as shown in Figure~\ref{fig:pretrain_data_pipeline}(E).
\end{itemize}

The resulting segmentation from this refined diarization process provides more accurate and consistently sized speaker turns compared to the baseline diarization output.

\textbf{Speech Transcription.} To obtain the language type and text transcription for each speech segment, we first apply the Whisper-large-v3 model~\citep{radford2023robust}\footnote{\url{https://huggingface.co/openai/whisper-large-v3}} to detect the spoken language type.
In this work, we retain only English and Mandarin segments for further transcription. 
For English segments, we directly use Whisper-large-v3 to generate both transcriptions and punctuation annotations. 
For Mandarin segments, we utilize the Paraformer-Zh~\citep{gao2022paraformer} model from the FunASR toolkit\footnote{\url{https://github.com/modelscope/FunASR}} to generate transcriptions along with character-level timestamps. Since Paraformer-Zh cannot output punctuation annotations, we add punctuation annotations with the following strategy: if the time gap between two consecutive characters is greater than $0.5$ seconds but less than $1.0$ second, we insert a ``comma''; if the gap exceeds $1.0$ second, we insert a ``period''.

\textbf{Implementation.} The data processing pipeline is deployed on a cluster of $30$ cloud instances. Each instance is equipped with $128$ virtual CPUs (vCores), $1$~TB of RAM, and $8$ NVIDIA L20 GPUs, powered by Intel Xeon Platinum 8575C processors that support vectorized acceleration instructions, including Advanced Matrix Extensions (AMX). 
In total, the cluster provides $3,840$ vCores, $30$~TB of memory, and $240$ NVIDIA L20 GPUs.
Following extensive optimization, the pipeline achieves a daily processing throughput of approximately $200,000$ hours of raw audio data. 

\subsection{SFT Data}
\label{sec:data/sftdata}
After the pre-training stage, we perform supervised fine-tuning (SFT) to enhance the performance of Kimi-Audio on instruction following and audio processing. The SFT data can be mainly categorized into three parts: audio understanding, speech conversation, and audio-to-text chat.

\begin{table*}[h!]
    \centering
    \caption{ List of datasets used for audio understanding and their training epoch in SFT stage.}
    \begin{threeparttable}
    \begin{tabular}{lrcc}
    \toprule
        Dataset & Audio Length (\#hours)  & Task Type & SFT Epochs\\ \midrule
        WenetSpeech~\citep{zhang2022wenetspeech} & $10,518$ & ASR & 2.0\\
        WenetSpeech4TTS~\citep{ma2024wenetspeech4tts} & $12,085$ & ASR & 2.0\\
        AISHELL-1~\citep{bu2017aishell} & $155$ & ASR & 2.0\\
        AISHELL-2~\citep{du2018aishell} & $1,036$ & ASR & 2.0\\
        AISHELL-3~\citep{shi2020aishell} & $65$ & ASR & 2.0\\
        Emilla~\citep{he2025emilia} & $98,305$ & ASR & 2.0\\
        Fleurs~\citep{conneau2023fleurs} & $17$ & ASR & 2.0\\
        CommonVoice~\citep{ardila2019common} & $43$ & ASR & 2.0\\
        KeSpeech~\citep{tang2021kespeech} & $1,428$ & ASR & 2.0\\
        Magicdata~\citep{yang2022open} & $747$ & ASR & 2.0\\
        zhvoice\textsuperscript{1} & $901$ & ASR & 2.0\\
        Libriheavy~\citep{kang2024libriheavy} & $51,448$ & ASR & 2.0\\
        MLS~\citep{Pratap2020MLSAL} & $45,042$ & ASR & 2.0\\
        Gigaspeech~\citep{chen2021gigaspeech} & $10,288$ & ASR & 2.0\\
        LibriSpeech~\citep{panayotov2015librispeech} & $960$ & ASR & 2.0\\
        CommonVoice~\citep{ardila2019common} & $1,854$ & ASR & 2.0\\
        Voxpopuli~\citep{wang2021voxpopuli} & $529$ & ASR & 2.0\\
        LibriTTS~\citep{zen2019libritts} & $568$ & ASR & 2.0\\
        CompA-R~\citep{ghosh2024gama} & $159$ & AQA & 2.0\\
        ClothoAQA~\citep{lipping2022clotho} & $7.4$ & AQA & $4.0$ \\
        AudioCaps~\citep{kim2019audiocaps} & $137$ & AAC & $2.0$ \\
        Clotho-v2~\citep{drossos2020clotho} & $24.0$ & AAC & $2.0$ \\
        MACS~\citep{martin2021ground} & $10.9$ & AAC & $2.0$ \\
        FSD50k~\citep{fonseca2021fsd50k} & $80.8$ & SEC & $2.0$ \\
        CochlScene~\citep{jeong2022cochlscene} & $169.0$ & ASC & $2.0$  \\
        Nonspeech7k~\citep{rashid2023nonspeech7k} & $6.2$ & SEC & $4.0$  \\
        MusicAVQA\textsubscript{audio-only}~\citep{li2022learning} & $77.1$ & AQA & $2.0$ \\
        WavCaps~\citep{mei2024wavcaps} & $3,793.3$ & AAC & $2.0$\\
        AVQA\textsubscript{audio-only}~\citep{yang2022avqa} & $112$ & AQA & $2.0$ \\
        IEMOCAP~\citep{tripathi2018multi} & $10$ & SER & $2.0$ \\
        MELD~\citep{poria2018meld} & $9$ & SER & $2.0$ \\
        RAVDESS~\citep{livingstone2018ryerson} & $3$ & SER & $2.0$ \\
        SAVEE~\citep{dataset} & $0.1$ & SER & $2.0$ \\
        ESD~\citep{zhou2022emotional} & $29$ & SER & $2.0$ \\
        TUT2016~\citep{Mesaros2016_EUSIPCO} & $10$ & ASC & $2.0$ \\
        TUT2017~\citep{Mesaros2016_EUSIPCO} & $13$ & ASC & $4.0$ \\
        TAU2022~\citep{heittola2022tau} & $67$ & ASC & $2.0$ \\
        ESC50~\citep{piczak2015dataset} & $1$ & SEC & $2.0$ \\
        VocalSound~\citep{gong_vocalsound} & $19$ & SEC & $4.0$ \\
        VGGSound~\citep{chen2020vggsound} & $513$ & SEC & $2.0$ \\
        UrbanSound8K~\citep{salamon2014dataset} & $9$ & SEC & $2.0$ \\
        FSD50K~\citep{fonseca2021fsd50k} & $74$ & SEC & $2.0$ \\
        Kimi Inhouse ASR Data & $55,000$ & ASR & $2.0$ \\
        Kimi Inhouse Audio Data & $5,200$ & AAC/AQA & $2.0$ \\
    \bottomrule
    \end{tabular}
    \end{threeparttable}
    \label{tab:sft_datasets_collected}
\end{table*}

\subsubsection{Audio Understanding}

\footnotetext[1]{\url{https://github.com/fighting41love/zhvoice}}

We mainly leverage open-source datasets for audio understanding. The collected datasets include $6$ tasks: Automatic Speech Recognition (ASR), Audio Question Answer (AQA), Automated Audio Caption (AAC), Speech Emotion Recognition (SER), Sound Event Classification (SEC), and Audio Scene Classification (ASC). The details of the datasets and the corresponding training epochs in the SFT stage are shown in Table~\ref{tab:sft_datasets_collected}. Besides the open-source datasets, we also utilize $55,000$ hours in-house ASR data and $5,200$ hours in-house audio data covering the AAC/AQA tasks.

\subsubsection{Speech Conversation}

To activate the Kimi-Audio model's ability to generate speech with diverse styles and high expressiveness in different conversation scenarios, we construct a large volume of speech conversation data, which consists of multi-turn conversations made up with a series of user queries and assistant responses. For user queries, we instruct LLMs to write the text of user queries and then convert them into speech with our Kimi-TTS system, where the prompt speech is randomly selected from a large timbre set containing more than $125$K timbres. For the assistant responses, we first select a voice actor as our Kimi-Audio speaker and synthesize the assistant responses with appropriate style and emotion with this single timbre. In the following, we introduce the data recording process for the Kimi-Audio speaker, as well as the Kimi-TTS and Kimi-VC systems used to synthesize assistant responses with diverse styles and expressiveness.

\textbf{Data Recording for Kimi-Audio Speaker.}
To achieve diverse and highly expressive styles and emotions in the generated speech, we select a voice actor as the Kimi-Audio speaker and meticulously record a dataset of this speaker in a professional recording studio. We pre-define over $20$ styles and emotions for recording, with each emotion further divided into $5$ levels to represent varying emotional intensities. For each style and emotional level, we record an audio as the reference to maintain the consistency of the emotion and style among different text sentences. The whole recording process is guided by a professional recording director.

\textbf{Kimi-TTS.} We develop a zero-shot text-to-speech synthesis (TTS) system, called Kimi-TTS, to generate speech with only a $3$-second prompt, while preserving the timbre, emotion, and style of the prompt speech. With the help of Kimi-TTS, we can synthesize speech for 1) the query text in diverse speakers/timbres with a large timbre set; 2) the response text with the styles and emotions recorded by the Kimi-Audio speaker, a voice actor selected by Kimi. Similar to the architecture of MoonCast~\citep{ju2025mooncast}, Kimi-TTS employs an LLM to generate speech tokens given the prompt speech and input text. Then a flow-matching-based speech detokenizer is used to generate high-quality speech waveforms. We train Kimi-TTS on about $1$M hours generated by the automatic data pipeline (Section~\ref{sec:data/pretraining}) and apply reinforcement learning to further enhance the robustness and quality of the generated speech.

\textbf{Kimi-VC.} Since it is difficult for the voice actor to record speech in any styles, emotions, and accents, we develop a voice conversion (VC) system, called Kimi-VC, to convert diverse and in-the-wild speech in different speakers/timbres into the timbre of Kimi-Audio speaker while preserving the styles, emotions, and accents. Built on the Seed-VC framework~\citep{liu2024zero}, Kimi-VC incorporates source timbre perturbation via a timbre-shifting model during training, which mitigates information leakage and ensures alignment between training and inference phases. To ensure high quality of voice conversion, we fine-tune the Kimi-VC model using speech data recorded by the Kimi-Audio speaker, a voice actor selected by Kimi.

\subsubsection{Audio-to-Text Chat}

\begin{table*}[h!]
    \centering
    \caption{ List of text dataset used in audio-to-text chat with their training epochs in SFT stage.}
    \begin{tabular}{lrcc}
    \toprule
        Dataset & \#Samples & SFT Epochs\\ \midrule
        Magpie-Pro~\citep{xu2024magpie}& $300$K & 2.0\\
        Magpie-MT~\citep{xu2024magpie} & $300$K & 2.0\\
        Evol-Instruct~\citep{surge2024openbezoar} & $143$K & 2.0\\
        Evol-Instruct-Code~\citep{surge2024openbezoar} & $80$K & 2.0\\
        Infinity-Instruct~\citep{InfinityInstruct2024} & $7$M & 2.0\\
        Synthia~\citep{tissera2023synthia} & $119$K & 2.0\\
        NuminaMath~\citep{li2024numinamath} & $860$K & 2.0\\
        Tulu3~\citep{lambert2024t} & $900$K & 2.0\\
        OpenHermes-2.5~\citep{OpenHermes2.5} & $1$M & 2.0\\
        OpenOrca~\citep{OpenOrca} & $2$M & 2.0\\
    \bottomrule
    \end{tabular}
    \label{tab:sft_datasets_synthesized}
\end{table*}

To help Kimi-Audio with the basic ability on chat, we collect open-source supervised fine-tuning data from text domain, as listed in Table~\ref{tab:sft_datasets_synthesized}, and then convert the user queries to speech with a variety of timbres, resulting in the audio-to-text chat data whose user query is speech while the assistant response is text. Considering that some text cannot be easily converted to speech, we perform several preprocessing steps on text by 1) filtering out text containing complex math, code, table, complex multilingual content, or too long content, 2) making colloquial rewriting, and 3) convert a single-turn question-answer data with complex instruction into multi-turn data with easy and concise instructions.

\section{Training}
\subsection{Pre-training}
\label{sec:training/pretraining}
The pre-training stage of Kimi-Audio aims to learn the knowledge from both the real-world audio and text domains and align them in the model's latent space, thereby facilitating complex tasks such as audio understanding, audio-to-text chat, and speech conversation. 
To this end, we design several pre-training tasks with the following aspects: 1) pre-training in the unimodality (i.e., audio and text) to learn the knowledge from each domain individually in Section~\ref{sec:training/pretraining/text_or_audio_pretraining}; 2) learning audio-text mapping in Section~\ref{sec:training/pretrain/asr}; 3) three audio-text interleaving tasks to further bridge two modalities in Section~\ref{sec:training/pretrain/interleave}. 

Formally, given a raw audio $A$, the data pre-processing pipeline (described in Section~\ref{sec:data/pretraining}) splits it into a series of segments $\{S_1, S_2, ..., S_N\}$, and each segment $S_i, i\in [1, N]$ consists of an audio $a_i$ and the corresponding transcription $t_i$. Furthermore, as illustrated in Section~\ref{sec:audio_tokenizer}, for an audio segment $a_i$, we extract both the continuous acoustic vectors $a_i^{c}$ and the discrete semantic tokens $a_i^d$.
To comply with the design of our model architecture in Section~\ref{sec:arc} which takes discrete semantic audio tokens as the main representation of the input and output, while adding continuous acoustic audio tokens in the input and discrete text tokens in the output, we denote the training sequence as $\{a^c_1/a^d_1/t_1, a^c_2/a^d_2/t_2, ..., a^c_N/a^d_N/t_N\}$, where $a^c_i/a^d_i/t_i$ denotes the semantic audio, acoustic audio, and text sequence for segment $i$. We make sure that the audio and text sequences have the same lengths by appending blank tokens to the shorter sequence. The actual pre-training segments can be either one or two of $a^c_i/a^d_i/t_i$, such as $a^d_i$, $t_i$, $a^c_i/a^d_i$, or $a^d_i/t_i$. For $a^c_i/a^d_i$, we add the continuous vectors $a_i^c$ and the semantic tokens $a_i^d$ (the semantic tokens will be converted into embedding using a lookup table) to obtain the final audio feature $a_i$. Thus, we use $a_i$ to represent $a^c_i/a^d_i$ for short. For $a^d_i/t_i$, we add the lookup embedding of the semantic tokens and text tokens as input and generate each token with its respective head, as described in Section~\ref{sec:arc}. 

With this notation, we formulate the following pre-training tasks in Table~\ref{tab:pretraining_tasks_definition} and introduce them as follows.

\subsubsection{Audio/Text Unimodal Pre-training}
\label{sec:training/pretraining/text_or_audio_pretraining}
We first learn the knowledge of text and audio separately. For text pre-training, we directly utilize the text data in MoonLight~\citep{liu2025muonscalablellmtraining}, which is high-quality and comprehensive for training large language models. We apply next-token prediction on text tokens solely. For audio pre-training, for each segment $S_i$, we apply next-token prediction on its discrete semantic token sequence $a^d_i$.

\begin{table}[t!]
    \centering
\caption{List of pre-training tasks. We design three categories pre-training tasks, including: 1) audio/text unimodal pre-training; 2) audio-text mapping pre-training; 3) audio-text interleaving pre-training. Notation: $a_i^d$ and $a_i^c$ denotes the discrete semantic tokens and continuous acoustic vectors for audio segment $i$ respectively, $a_i$ denotes the combination of $a_i^d$ and $a_i^c$ for audio segment $i$, underline means it will receive loss during training.}
    \begin{tabular}{lllc}
    \toprule
Category  & Pre-training Task              & Task Formulation            & Task Weight \\ \midrule
\multirow{2}{*}{Audio/Text Unimodal}  &Text Only & $\underline{t_1}, \underline{t_2}, ..., \underline{t_N}    $    &  $7$ \\
&Audio Only & $\underline{a_1^d}, \underline{a_2^d}, ... , \underline{a_N^d}$ &    $1$ \\ \midrule
\multirow{2}{*}{Audio-Text Mapping} &Audio to Text & $a_1, \underline{t_1}, a_2, \underline{t_2}, ..., a_N, \underline{t_N}$ & $1$ \\
&Text to Audio & $t_1, \underline{a_1^d}, t_2, \underline{a_2^d}, ..., t_N, \underline{a_N^d}$ & $1$ \\ \midrule
\multirow{3}{*}{Audio-Text Interleaving} &Audio to Semantic &   $a_1, \underline{a_2^d}, a_3, \underline{a_4^d}, ..., a_{N-1}, \underline{a_N^d}$ & $1$ \\
&Audio to Text & $a_1, \underline{t_2}, a_3, \underline{t_4}, ..., a_{N-1}, \underline{t_N}$ & $1$ \\
&Audio to Semantic and Text & $a_1, \underline{a_2^d} /\underline{t_2}, a_3, \underline{a_4^d}/\underline{t_4}, ..., a_{N-1}, \underline{a_N^d}/ \underline{t_N}$ & $2$ \\
    \bottomrule
\end{tabular}
    \label{tab:pretraining_tasks_definition}
\end{table}

\subsubsection{Audio-Text Mapping Pre-training}
\label{sec:training/pretrain/asr}
Intuitively, in order to align audio and text in a unified space, it is helpful to learn a mapping between two modalities. Thus, we design the automatic speech recognition (ASR) and text-to-speech synthesis (TTS) pre-training tasks. For ASR, we formulate the training sequence as $\{a_1, t_1, a_2, t_2, ..., a_N, t_N\}$. For TTS, we formulate the training sequence as $\{t_1, a_1^d,t_2, a_2^d, ..., t_N, a_N^d\}$. We only calculate the loss on text tokens for ASR and on audio semantic tokens for TTS.

\subsubsection{Audio-Text Interleaving Pre-training}
\label{sec:training/pretrain/interleave}
To further bridge the gap between audio and text modalities, we design three audio-text interleaving pre-training tasks.

\begin{itemize}[leftmargin=*]
    \item \textbf{Audio to semantic token interleaving.} We formulate the training sequence as $\{a_1, a_2^d, a_3, a_4^d, ..., a_{N-1}, a_N^d\}$\footnote{It is also possible that the first segment is $a_1^d$, or the last segment is $a_{N}$.}. Then we only calculate the loss on the semantic audio tokens $a_i^d$, but not on $a_{i-1}$.
    \item \textbf{Audio to text interleaving.} We formulate the training sequence as $\{a_1, t_2, a_3, t_4, ..., a_{N-1}, t_N\}$. We only calculate the loss on text tokens $t_i$.
    \item \textbf{Audio to semantic token + text interleaving.} We formulate the training sequence as $\{a_1, a_2^d/t_2, a_3, a_4^d/t_4, ..., a_{N-1}, a_N^d/t_N\}$. For $a_i^d/t_i$, as the semantic audio token sequence is always longer than the text token sequence, the prediction of the semantic token is like a streaming text-to-speech task as in Section~\ref{sec:training/pretrain/asr}. Empirically, we find that the prediction of the first few semantic tokens is hard because the model needs to concurrently predict the next text token and its semantic audio token. We address this issue by delaying the prediction of the first several semantic audio tokens by prepending $6$ special blank tokens ($6$ is determined by trading off the generation quality and latency according to preliminary experiments) to the semantic audio tokens.
\end{itemize}

\subsubsection{Pre-training Recipe}
We initialize the audio LLM of Kimi-Audio from the pre-trained Qwen2.5 7B model~\citep{qwen2.5} and extend its vocabulary with semantic audio tokens and special tokens. We perform pre-training on the above pre-training tasks with the corresponding task weight 
$1:7:1:1:1:1:2$, as shown in Table~\ref{tab:pretraining_tasks_definition}. We pre-train Kimi-Audio using $585$B audio tokens and $585$B text tokens with $1$ epoch. We use AdamW~\citep{loshchilov2017decoupled} optimizer with a learning rate schedule from $2e^{-5}$ to $2e^{-6}$ in a cosine decay. We use $1$\% tokens for learning rate warmup.

The continuous acoustic feature extraction module in audio tokenizer is initialized from Whisper large-v3~\citep{radford2023robust}, which can capture the fine-grained acoustic characteristics inherent in the input audio signal. During the initial phases of model pretraining (about $20$\% tokens in pretraining), the parameters of this whisper-based feature extractor are kept frozen. Subsequently, the feature extractor is unfrozen, enabling its parameters to be fine-tuned jointly with the rest of the model, allowing it to adapt more specifically to the nuances of the training data and the requirements of the target tasks.

\subsection{Supervised Fine-tuning}
\label{sec:training/sft}

\subsubsection{Formulation}
After pre-training Kimi-Audio with massive real-world audio and text data, we perform supervised fine-tuning to equip it with the ability of instruction-following. We have the following design choices: 1) considering the downstream tasks are diverse, we do not set special task switching operations, but use natural language as instructions for each task; 2) for instruction, we construct both the audio and text versions (i.e., the audio is generated by Kimi-TTS in a zero-shot way given the text) and randomly choose one during training; 3) to enhance the robustness of instruction-following capability, we construct $200$ instructions for ASR task and $30$ instructions for other tasks by LLM and randomly choose one for each training sample. As described in Section \ref{sec:data/sftdata}, we build about $300$K hours of data for supervised fine-tuning.

\subsubsection{Fine-tuning Recipe}
As shown in Table \ref{tab:sft_datasets_collected} and Table \ref{tab:sft_datasets_synthesized}, we fine-tune Kimi-Audio on each data source with $2$-$4$ epochs based on comprehensive ablation experiments. We use AdamW~\citep{loshchilov2017decoupled} optimizer with a learning rate schedule from $1e^{-5}$ to $1e^{-6}$ in a cosine decay. We use $10$\% tokens for learning rate warmup.

\subsection{Training of Audio Detokenizer}

We train the audio detokenizer in three stages. Firstly, we use about $1$M hours of audio from the pre-training data described in Section~\ref{sec:data/pretraining} and pre-train both the flow-matching model and the vocoder to learn audio with diverse timbre, prosody, and quality. Secondly, we adopt the chunk-wise fine-tuning strategy with a dynamic chunk size from $0.5$ seconds to $3$ seconds on the same pre-training data~\citep{ju2025mooncast}. Finally, we fine-tune on the high-quality single-speaker recording data from the Kimi-Audio speaker.

\section{Inference and Deployment}
\label{sec:deploy}
Kimi-Audio is designed to handle various audio-related tasks, such as speech recognition, audio understanding, audio-to-text chat, and speech-to-speech conversation. We take real-time speech-to-speech conversation as an example to illustrate the practices in Kimi-Audio deployment, since this task is more complicated than the rest of audio tasks in terms of infrastructure and engineering efforts. We first introduce the workflow of real-time speech conversation between the client (e.g., Kimi APP or web browser) and the server (Kimi-Audio service) and then describe the practices of product deployment.

\subsection{Workflow of Real-Time Speech Conversation}
\label{sec:deploy_workflow}

\begin{figure}
    \centering
    \includegraphics[width=0.4\columnwidth, height=0.5\textheight, trim=0.6cm 1.0cm 0.6cm 1.0cm, clip=true]
    {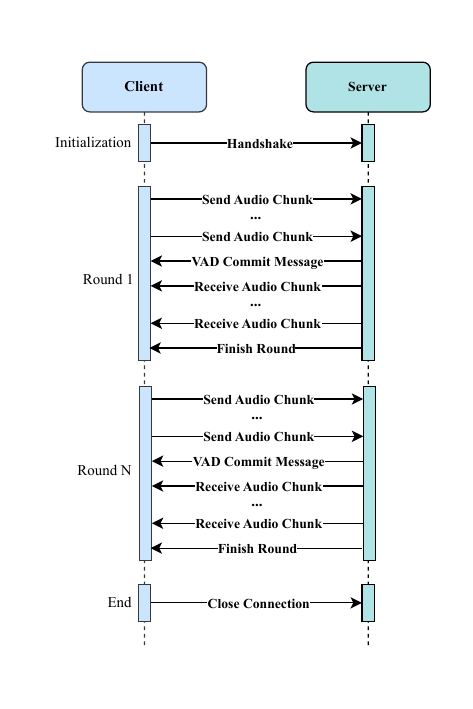}
    \caption{The client-server communication for real-time speech-to-speech conversation in Kimi-Audio.}
    \label{fig:deploy_workflow}
\end{figure}

The workflow of a speech-to-speech conversation between the user client (e.g., Kimi APP) and the server (Kimi-Audio service) is illustrated in Figure~\ref{fig:deploy_workflow}. The workflow proceeds in the following manner for each conversation round:

\begin{itemize} 
\item The user speaks to the client (e.g., Kimi APP or web browser), and the audio data is collected and streamed to the server. 
\item On the server side, a voice activity detection (VAD) module determines if the user has finished speaking. 
\item Once the user stops speaking, the server sends a commit signal and initiates the inference process of the Kimi-Audio model. 
\item During inference, the client receives audio chunks in real-time as they are generated and starts playing them for the user. 
\item The client (mobile phone or web browser) plays the received audio chunks back to the user. 
\end{itemize}

The inference process of Kimi-Audio on the server side for each round follows these steps. First, the input audio is converted to discrete semantic tokens and continuous acoustic vectors using the audio tokenizer. Next, the input to the Audio LLM is assembled by concatenating the system prompt tokens, audio tokens, and conversation history tokens. The token sequence is then passed to the Audio LLM, which generates output tokens. Finally, the output tokens are converted back into an audio waveform using the detokenizer.

\begin{figure}[t]
    \centering
\includegraphics[width=1.0\columnwidth,trim=0.5cm 12.0cm 0.5cm 0.5cm,clip=true]
    {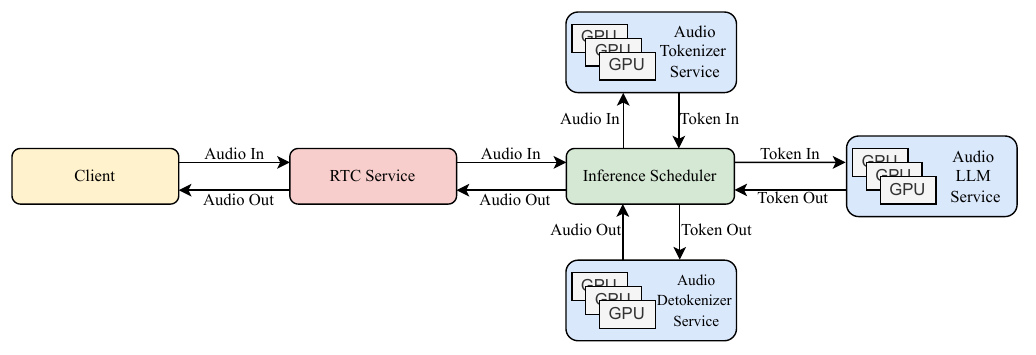}
    \caption{The workflow of production deployment for real-time speech-to-speech conversation in Kimi-Audio.}
    \label{fig:deploy_workflow_product}
\end{figure}

\subsection{Production Deployment}

As shown in Figure~\ref{fig:deploy_workflow_product}, in a production environment, all core components: Audio Tokenizer, Audio LLM, and Audio Detokenizer, are computationally intensive, requiring a scalable and efficient infrastructure. To address this, we designed the production deployment architecture as follows.

\textbf{Kimi-Audio RTC Service.} This service interfaces with the client, receiving audio from the user, forwarding it to the Inference Scheduler, and returning the generated audio chunks to the client. We use the WebRTC protocol to ensure a stable and low-latency connection.

\textbf{Inference Scheduler.} The Inference Scheduler manages the conversation flow by maintaining conversation history as tokens in a storage backend. For each interaction round, it performs the following steps:
\begin{itemize} 
\item Call the Tokenizer Service to convert the user’s audio into tokens.

\item Construct the model input by combining the new tokens with the conversation history.

\item Send the input to the LLM Service to generate response tokens.

\item Call the Detokenizer Service to convert the response tokens into audio output.
\end{itemize}
Additionally, it stores all output tokens as part of the ongoing conversation history to ensure continuity in the dialogue.

\textbf{Tokenizer/Detokenizer/LLM Services}: These services handle model inference and are equipped with a load balancer and multiple inference instances to handle requests in parallel, ensuring scalability.

This modular architecture ensures that Kimi-Audio can scale effectively to meet the performance demands of real-time speech interactions while maintaining low latency and high availability in production.

\section{Evaluation}
Evaluating audio foundation models and comparing with previous state-of-the-art systems are challenging, due to some inherent issues in the audio community. Thus, we first develop a fair, reproducible, and comprehensive evaluation toolkit for audio foundation models in Section~\ref{sec:eval_tool}, and then evaluate Kimi-Audio on a variety of audio processing tasks including speech recognition, general audio understanding, audio-to-text chat, and speech conversation, and compare Kimi-Audio with previous systems to demonstrate the advantages in Section~\ref{sec:eval_result}. 
\subsection{Evaluation Toolkit} 
\label{sec:eval_tool}
Even if an audio foundation model is fully open-source, it is still troublesome to reproduce the same results as reported in its paper or technical report, let alone those closed-source models. We analyze the challenges in evaluating and comparing audio foundation models in various audio processing tasks as follows:
\begin{itemize}
    \item \textbf{Limitations in Metrics.} Current practices suffer from inconsistent metric implementations (e.g., variations in Word Error Rate calculation due to different text normalizations) and inadequate assessment methods (e.g., relying solely on exact string matching for tasks like audio question answering fails to capture the semantic correctness of complex LLM responses).
    \item \textbf{Diverse Configurations.} Reproducibility is severely hampered by the high sensitivity of model performance to inference parameters such as decoding temperature, system prompts, and task prompts.
    \item \textbf{Lack of Generation Evaluation} While progress has been made in understanding tasks, assessing the quality and coherence of the generated audio response still lacks benchmarks.
\end{itemize}

To address these critical limitations, we develop an open-source evaluation toolkit for audio foundation models on audio understanding, generation, and conversation tasks. It currently integrates and supports Kimi-Audio and a series of recent audio LLMs~\citep{chu2024qwen2,xu2025qwen25omnitechnicalreport,zeng2024glm,li2025baichuan,huang2025step}, and can be leveraged to evaluate any other audio foundation models. The toolkit provides the following features and benefits:
\begin{itemize}
\item We implement a standardized WER calculation (based on Qwen-2-Audio~\citep{chu2024qwen2}) and integrate GPT-4o-mini as an intelligent judge (following \citep{chen2024voicebench}) for tasks like audio question answering. This approach overcomes the limitations of inconsistent metrics and simplistic string matching, enabling fair comparison.
\item Our toolkit offers a single unified platform supporting diverse models and versions, simplifying side-by-side comparisons. It provides a crucial structure for defining and sharing standardized inference parameters and prompting strategies (``recipes''), thereby directly addressing inconsistencies in evaluation setups and fostering greater reproducibility across different research works.
\item We record and release an evaluation benchmark to test the ability of audio LLMs on speech conversation from the perspective of 1) speech control on emotion, speed, and accent; 2) empathy conversation; and 3) diverse styles such as storytelling and tongue twister.
\end{itemize}

We open-source this toolkit to the community \url{https://github.com/MoonshotAI/Kimi-Audio-Evalkit}. We believe this toolkit can serve as a valuable asset to advance the field by promoting more reliable and comparable benchmarking. We actively encourage researchers and developers to utilize it, contribute by adding new models and datasets, and help refine standardized evaluation protocols and inference recipes. In this way, we can build a better ecosystem for the audio community.

\begin{table*}[t]
\centering
\caption{Performance of Kimi-Audio and baseline models on ASR task. Best results are in bold.}
\label{tab:asr_performance}
\begin{tabular}{c l l}
\toprule
\textbf{Datasets} & \textbf{Model} & \textbf{Performance (WER$\downarrow$)} \\
\midrule
\multirow{5}{*}{\begin{tabular}[c]{@{}c@{}}\textbf{LibriSpeech}~\citep{panayotov2015librispeech} \\ test-clean | test-other\end{tabular}} & Qwen2-Audio-base &  1.74 | 4.04 \\ 
& Baichuan-Audio-base & 3.02 | 6.04 \\
& Step-Audio-chat & 3.19 | 10.67 \\
 & Qwen2.5-Omni & 2.37 | 4.21 \\
 \cmidrule{2-3}
 & Kimi-Audio & \textbf{1.28} | \textbf{2.42} \\
\midrule
\multirow{5}{*}{\begin{tabular}[c]{@{}c@{}}\textbf{Fleurs}~\citep{conneau2023fleurs} \\ zh | en\end{tabular}} & Qwen2-Audio-base &  3.63 | 5.20 \\ 
& Baichuan-Audio-base & 4.15 | 8.07 \\
& Step-Audio-chat & 4.26 | 8.56 \\
 & Qwen2.5-Omni & 2.92 | \textbf{4.17} \\
 \cmidrule{2-3}
 & Kimi-Audio & \textbf{2.69} | 4.44 \\
\midrule
\multirow{5}{*}{\begin{tabular}[c]{@{}c@{}}\textbf{AISHELL-1}~\citep{bu2017aishell} \end{tabular}}  & Qwen2-Audio-base &  1.52 \\ 
& Baichuan-Audio-base & 1.93  \\
& Step-Audio-chat & 2.14 \\
 & Qwen2.5-Omni & 1.13  \\
 \cmidrule{2-3}
 & Kimi-Audio & \textbf{0.60} \\
\midrule
\multirow{5}{*}{\begin{tabular}[c]{@{}c@{}}\textbf{AISHELL-2}~\citep{du2018aishell} ios \end{tabular}}  & Qwen2-Audio-base &  3.08 \\ 
& Baichuan-Audio-base & 3.87  \\
& Step-Audio-chat & 3.89 \\
 & Qwen2.5-Omni & \textbf{2.56} \\
 \cmidrule{2-3}
 & Kimi-Audio & \textbf{2.56} \\
\midrule
\multirow{5}{*}{\begin{tabular}[c]{@{}c@{}}\textbf{WenetSpeech}~\citep{zhang2022wenetspeech} \\ test-meeting | test-net \end{tabular}} & Qwen2-Audio-base &   8.40 | 7.64 \\ 
& Baichuan-Audio-base & 13.28 | 10.13 \\
& Step-Audio-chat & 10.83 | 9.47 \\
 & Qwen2.5-Omni & 7.71 | 6.04 \\
 \cmidrule{2-3}
 & Kimi-Audio & \textbf{6.28} | \textbf{5.37} \\
\midrule
\multirow{5}{*}{\begin{tabular}[c]{@{}c@{}}\textbf{Kimi-ASR Internal Testset} \\ subset1 | subset2 \end{tabular}} & Qwen2-Audio-base &  2.31 | 3.24 \\ 
& Baichuan-Audio-base & 3.41 | 5.60  \\
& Step-Audio-chat &  2.82 | 4.74 \\
 & Qwen2.5-Omni & 1.53 | 2.68 \\
 \cmidrule{2-3}
 & Kimi-Audio & \textbf{1.42} | \textbf{2.44} \\
\bottomrule
\end{tabular}
\end{table*}

\subsection{Evaluation Results}
\label{sec:eval_result}
In this section, based on our evaluation toolkit, we detail the evaluation of Kimi-Audio across a comprehensive suite of audio processing tasks, including Automatic Speech Recognition (ASR), audio understanding, audio-to-text chat, and speech conversation. We compare Kimi-Audio against other audio foundation models (Qwen2-Audio~\citep{chu2024qwen2}, Baichuan-Audio~\citep{li2025baichuan}, Step-Audio~\citep{huang2025step}, GLM-4-Voice~\citep{zeng2024glm}, and Qwen2.5-Omini~\citep{xu2025qwen2}) using established benchmarks and internal test sets.

\begin{table*}[t]
\centering
\caption{Performance of Kimi-Audio and baseline models on audio understanding task. Best results are in bold.}
\label{tab:audio_understanding_performance}
\begin{tabular}{c l l}
\toprule
\textbf{Datasets} & \textbf{Model} & \textbf{Performance$\uparrow$} \\
\midrule
\multirow{6}{*}{\begin{tabular}[c]{@{}c@{}}\textbf{MMAU}~\citep{sakshi2024mmau} \\ music | sound | speech \end{tabular}} & Qwen2-Audio-base &  58.98 | 69.07 | 52.55 \\ 
& Baichuan-chat & 49.10 | 59.46 | 42.47 \\
 & GLM-4-Voice &  38.92 | 43.54 | 32.43 \\
& Step-Audio-chat & 49.40 | 53.75 | 47.75 \\
 & Qwen2.5-Omni & \textbf{62.16} | 67.57 | 53.92 \\
  \cmidrule{2-3}
 & Kimi-Audio & 61.68 | \textbf{73.27} | \textbf{60.66} \\
\midrule
\multirow{5}{*}{\begin{tabular}[c]{@{}c@{}}\textbf{ClothoAQA}~\citep{lipping2022clotho} \\ test | dev \end{tabular}} & Qwen2-Audio-base & 71.73 | 72.63  \\ 
& Baichuan-chat & 48.02 | 48.16 \\
& Step-Audio-chat & 45.84 | 44.98 \\
 & Qwen2.5-Omni & \textbf{72.86} | 73.12 \\
  \cmidrule{2-3}
 & Kimi-Audio & 71.24 | \textbf{73.18} \\
\midrule
\multirow{5}{*}{\begin{tabular}[c]{@{}c@{}}\textbf{VocalSound}~\citep{gong_vocalsound} \end{tabular}} & Qwen2-Audio-base & 93.82  \\ 
& Baichuan-Audio-base & 58.17 \\
& Step-Audio-chat & 28.58 \\
 & Qwen2.5-Omni & 93.73 \\
  \cmidrule{2-3}
 & Kimi-Audio & \textbf{94.85} \\
\midrule
\multirow{5}{*}{\begin{tabular}[c]{@{}c@{}}\textbf{Nonspeech7k}~\citep{rashid2023nonspeech7k} \end{tabular}}  & Qwen2-Audio-base &  87.17 \\ 
& Baichuan-chat & 59.03  \\
& Step-Audio-chat & 21.38 \\
 & Qwen2.5-Omni & 69.89  \\
  \cmidrule{2-3}
 & Kimi-Audio & \textbf{93.93} \\
\midrule
\multirow{5}{*}{\begin{tabular}[c]{@{}c@{}}\textbf{MELD}~\citep{poria2018meld} \end{tabular}}  & Qwen2-Audio-base &  51.23 \\ 
& Baichuan-chat  & 23.59 \\
& Step-Audio-chat & 33.54 \\
 & Qwen2.5-Omni & 49.83 \\
  \cmidrule{2-3}
 & Kimi-Audio & \textbf{59.13} \\
\midrule
\multirow{5}{*}{\begin{tabular}[c]{@{}c@{}}\textbf{TUT2017}~\citep{Mesaros2016_EUSIPCO} \end{tabular}}  & Qwen2-Audio-base &  33.83 \\ 
& Baichuan-Audio-base & 27.9 \\
& Step-Audio-chat & 7.41 \\
 & Qwen2.5-Omni & 43.27 \\
  \cmidrule{2-3}
 & Kimi-Audio & \textbf{65.25} \\
\midrule
\multirow{5}{*}{\begin{tabular}[c]{@{}c@{}}\textbf{CochlScene}~\citep{jeong2022cochlscene}  \\ test | dev \end{tabular}} & Qwen2-Audio-base &  52.69 | 50.96 \\ 
& Baichuan-Audio-base & 34.93 | 34.56 \\
& Step-Audio-chat & 10.06 | 10.42 \\
 & Qwen2.5-Omni & 63.82 | 63.82 \\
  \cmidrule{2-3}
 & Kimi-Audio & \textbf{79.84} | \textbf{80.99} \\
\bottomrule
\end{tabular}
\end{table*}

\subsubsection{Automatic Speech Recognition}
The ASR capabilities of Kimi-Audio were evaluated on diverse datasets spanning multiple languages and acoustic conditions. As presented in Table~\ref{tab:asr_performance}, Kimi-Audio consistently demonstrates superior performance compared to previous models. We report Word Error Rate (WER) on these datasets, where lower values indicate better performance.

Notably, Kimi-Audio achieves the best results on the widely-used LibriSpeech~\citep{panayotov2015librispeech} benchmark, attaining error rates of $1.28$ on test-clean and $2.42$ on test-other, significantly outperforming models like Qwen2-Audio-base and Qwen2.5-Omni.
For Mandarin ASR benchmarks, Kimi-Audio sets SOTA results on AISHELL-1~\citep{bu2017aishell} ($0.60$) and AISHELL-2 ios~\citep{du2018aishell} ($2.56$). Furthermore, it excels on the challenging WenetSpeech~\citep{zhang2022wenetspeech} dataset, achieving the lowest error rates on both test-meeting and test-net. Finally, evaluation on our internal Kimi-ASR test set confirms the model robustness. These results demonstrate the strong ASR capabilities of Kimi-Audio across various domains and languages.

\subsubsection{Audio Understanding}
Beyond speech recognition, we assess Kimi-Audio's ability to comprehend diverse audio signals, including music, sound events, and speech. Table~\ref{tab:audio_understanding_performance} summarizes the performance on various audio understanding benchmarks, where higher scores generally indicate better performance.

On the MMAU benchmark~\citep{sakshi2024mmau}, Kimi-Audio demonstrates superior understanding across sound category ($73.27$), and speech category ($60.66$). Similarly, it outperforms other models on the MELD~\citep{poria2018meld} speech emotion understanding task, scoring $59.13$. Kimi-Audio also leads on tasks involving non-speech sound classification (VocalSound~\citep{gong_vocalsound} and Nonspeech7k~\citep{rashid2023nonspeech7k}) and acoustic scene classification (TUT2017~\citep{Mesaros2016_EUSIPCO} and CochlScene~\citep{jeong2022cochlscene}). These results highlight Kimi-Audio's advanced capabilities in interpreting complex acoustic information beyond simple speech recognition.

\begin{table*}[t]
\centering
\caption{Performance of Kimi-Audio and baseline models on the tasks of audio-to-text chat. Best results are in bold.}
\label{tab:audio_text_performance}
\begin{tabular}{c l l}
\toprule
\textbf{Datasets} & \textbf{Model} & \textbf{Performance$\uparrow$} \\
\midrule
\multirow{6}{*}{\begin{tabular}[c]{@{}c@{}}\textbf{OpenAudioBench}~\citep{li2025baichuan} \\ AlpacaEval | Llama Questions | \\ Reasoning QA | TriviaQA | Web Questions \end{tabular}} & Qwen2-Audio-chat &  57.19 | 69.67 | 42.77 | 40.30 | 45.20 \\ 
& Baichuan-chat & 59.65 | 74.33 | 46.73 | 55.40 | 58.70 \\
& GLM-4-Voice & 57.89 | 76.00 | 47.43 | 51.80 | 55.40 \\
& Step-Audio-chat & 56.53 | 72.33 | 60.00 | 56.80 | \textbf{73.00} \\
 & Qwen2.5-Omni & 72.76 | 75.33 | \textbf{63.76} | 57.06 | 62.80  \\
  \cmidrule{2-3}
 & Kimi-Audio & \textbf{75.73} | \textbf{79.33} | 58.02 | \textbf{62.10} | 70.20 \\
\midrule
\multirow{6}{*}{\begin{tabular}[c]{@{}c@{}}\textbf{VoiceBench}~\citep{chen2024voicebench} \\ AlpacaEval | CommonEval | \\ SD-QA | MMSU\end{tabular}}  & Qwen2-Audio-chat &  3.69 | 3.40 | 35.35 | 35.43 \\ 
& Baichuan-chat & 4.00 | 3.39 | 49.64 | 48.80 \\
& GLM-4-Voice & 4.06 | 3.48 | 43.31 | 40.11 \\
& Step-Audio-chat & 3.99 | 2.99 | 46.84 | 28.72 \\
 & Qwen2.5-Omni & 4.33 | 3.84 | 57.41 | 56.38 \\
  \cmidrule{2-3}
 & Kimi-Audio & \textbf{4.46} | \textbf{3.97} | \textbf{63.12} | \textbf{62.17} \\
\midrule
\multirow{6}{*}{\begin{tabular}[c]{@{}c@{}}\textbf{VoiceBench}~\citep{chen2024voicebench} \\ OpenBookQA | IFEval | \\ AdvBench | Avg \end{tabular}}  & Qwen2-Audio-chat &  49.01 | 22.57 | 98.85 | 54.72 \\ 
& Baichuan-chat & 63.30 | 41.32 | 86.73 | 62.51 \\
& GLM-4-Voice & 52.97 | 24.91 | 88.08 | 57.17 \\
& Step-Audio-chat & 31.87 | 29.19 | 65.77 |  48.86 \\
 & Qwen2.5-Omni &  79.12 | 53.88 | 99.62 | 72.83 \\
  \cmidrule{2-3}
 & Kimi-Audio & \textbf{83.52}  | \textbf{61.10} | \textbf{100.00} | \textbf{76.93} \\
\bottomrule
\end{tabular}
\end{table*}

\subsubsection{Audio-to-Text Chat}
We evaluate the ability of Kimi-Audio to engage in text conversations based on audio input using the OpenAudioBench~\citep{li2025baichuan} and VoiceBench benchmarks~\citep{chen2024voicebench}. These benchmarks assess various aspects like instruction following, question answering, and reasoning. Performance metrics are benchmark-specific, with higher scores indicating better conversational ability. The results are presented in Table~\ref{tab:audio_text_performance}.

On OpenAudioBench, Kimi-Audio achieves state-of-the-art performance on several sub-tasks, including AlpacaEval, Llama Questions, and TriviaQA, and achieves highly competitive performance on Reasoning QA and Web Questions.

The VoiceBench evaluation further confirms Kimi-Audio's strengths. It consistently outperforms all compared models on AlpacaEval ($4.46$), CommonEval ($3.97$), SD-QA ($63.12$), MMSU ($62.17$), OpenBookQA ($83.52$), Advbench ($100.00$), and IFEval ($61.10$). Kimi-Audio's overall performance across these comprehensive benchmarks demonstrates its superior ability in audio-based conversation and complex reasoning tasks.

\subsubsection{Speech Conversation}
Finally, we assess the end-to-end speech conversation capabilities of Kimi-Audio based on subjective evaluations across multiple dimensions. As shown in Table~\ref{tab:smo_performance}, Kimi-Audio was compared against models like GPT-4o and GLM-4-Voice based on human ratings (on a $1$-$5$ scale, higher is better).

Excluding GPT-4o, Kimi-Audio achieves the highest scores for emotion control, empathy, and speed control. While GLM-4-Voice shows slightly better accent control, Kimi-Audio achieves a strong overall average score of $3.90$. This score is higher than Step-Audio-chat ($3.33$), GPT-4o-mini ($3.45$), and GLM-4-Voice ($3.65$), and remains a small margin with GPT-4o ($4.06$). Overall, the evaluation results demonstrate Kimi-Audio's proficiency in generating expressive and controllable speech.

\begin{table*}[t]
\centering
\caption{Performance of Kimi-Audio and baseline models on speech conversation. Best results are in bold and second-best results are underlined.}
\label{tab:smo_performance}
\begin{tabular}{l c c c c c c}
\toprule
\textbf{Model} & Speed Control & Accent Control & Emotion Control & Empathy & Style Control & Avg \\
\midrule
GPT-4o & \underline{4.21} & \textbf{3.65} & 4.05 & \textbf{3.87} & \textbf{4.54} & \textbf{4.06} \\
Step-Audio-chat & 3.25 & 2.87 & 3.33 & 3.05 & \underline{4.14} & 3.33 \\
GLM-4-Voice & 3.83 & 3.51 & 3.77 & 3.07 & 4.04 & 3.65 \\
GPT-4o-mini & 3.15 & 2.71 & \underline{4.24} & 3.16 & 4.01 & 3.45 \\
\midrule
Kimi-Audio & \textbf{4.30} & \underline{3.45} & \textbf{4.27} & \underline{3.39} & 4.09 & \underline{3.90} \\
\bottomrule
\end{tabular}
\end{table*}

\section{Related Work}
The application of large language models (LLMs) to audio tasks has led to remarkable progress across a wide range of domains, including automatic speech recognition (ASR), audio understanding, text-to-speech synthesis (TTS), general audio generation, and speech-based human-computer interaction. These efforts explore how to bridge the gap between raw acoustic signals and linguistic reasoning by treating audio as a tokenizable sequence, enabling LLMs to process or generate audio in a language-like manner.

\paragraph{ASR and Audio Understanding}
A number of LLM-based systems have been developed to improve automatic speech recognition (ASR) and broader audio understanding tasks. Whisper~\citep{radford2023robust} serves as a powerful audio encoder, and when combined with large language models (LLMs), it significantly enhances the performance of speech understanding systems. This approach has been successfully utilized in models such as Qwen-Audio~\citep{chu2023qwen}, Qwen2-Audio~\citep{chu2024qwen2}, SALMONN~\citep{tang2023salmonn}, and OSUM~\citep{geng2025osum} These systems, however, are mostly limited to understanding tasks and do not natively support audio output.

\paragraph{TTS and Audio Generation}
For speech synthesis and general audio generation, models such as AudioLM~\citep{borsos2023audiolm}, VALL-E~\citep{wang2023neural}, and LLASA~\citep{ye2025llasa} tokenize audio via neural codecs and use decoder-only language models for autoregressive generation. Other efforts like UniAudio~\citep{yang2023uniaudio} and VoiceBox~\citep{le2023voicebox} extend these methods with hybrid tokenization or flow matching to improve quality and control. While these models can produce high-fidelity audio, they typically focus on generation only and lack understanding and conversation capabilities or instruction-following speech interaction.

\paragraph{Speech Conversation and Real-Time Dialogue}
Recent models have moved toward enabling real-time, end-to-end speech interaction. Moshi~\citep{defossez2024moshi}, GLM-4-Voice~\citep{zeng2024glm}, and Mini-Omni~\citep{xie2024mini} adopt interleaved or parallel decoding to support simultaneous generation of text and audio tokens, facilitating low-latency dialogue systems. OmniFlatten~\citep{zhang2024omniflatten} introduces a progressive training pipeline to adapt a frozen LLM for full-duplex conversation. LLaMA-Omni~\citep{fang2024llama} and Freeze-Omni~\citep{wang2024freeze} further refine duplex speech interaction through streaming decoders or multi-task alignment strategies. However, these systems often rely heavily on speech-only datasets and compromise language modeling quality or generality due to limited pre-training.

\paragraph{Toward Universal Audio-Language Foundation Models}
A small number of recent works aim to unify understanding and generation within a single multimodal model. Baichuan-Audio~\citep{li2025baichuan} uses a multi-codebook discretization to capture both semantic and acoustic features, enabling real-time interaction and strong question-answering capabilities. However, its focus on speech domain limits its broader applicability, especially for non-speech audio tasks like music or environmental sound. Step-Audio~\citep{huang2025step}, on the other hand, provides a powerful solution for real-time speech interaction with a 130B-parameter unified speech-text multimodal model. While Step-Audio demonstrates strong performance, its dependency on synthetic voice data generation and the high computational costs associated with its 130B parameters pose significant barriers to accessibility and cost-effectiveness for a broader user base. Qwen2.5-Omni~\citep{xu2025qwen25omnitechnicalreport} introduces a Thinker-Talker architecture for simultaneous text and speech decoding and achieves strong benchmark results, but its design primarily emphasizes streaming inference, and it lacks an extensive pre-training phase on raw audio.

\paragraph{Kimi-Audio}
Kimi-Audio advances beyond these limitations by introducing a truly universal and open-source audio foundation model that supports speech recognition, audio understanding, audio generation, and speech conversation in a single framework. It adopts a hybrid input representation combining Whisper-derived continuous acoustic features and discrete semantic tokens ($12.5$Hz), ensuring rich perception and efficient modeling. The audio LLM is initialized from a text LLM, with dual-generation heads for text and audio, and a chunk-wise flow-matching detokenizer paired with BigVGAN~\citep{lee2022bigvgan} to produce expressive and low-latency speech.

Most critically, Kimi-Audio features extensive multimodal pretraining on $13$ million hours of curated audio data across speech, music, and environmental sound—a scale far exceeding prior works. The pretraining tasks include audio-only, text-only, audio-to-text, and interleaved modalities, which enable the model to learn generalizable audio reasoning and maintain strong language abilities. This is followed by instruction-based fine-tuning across diverse tasks, leading to state-of-the-art results on ASR, general audio understanding, audio-text chat, and speech conversation benchmarks.

In contrast to existing models that are either limited in scope, lack pre-training, or are not publicly available, Kimi-Audio is a fully open-source, pre-trained, instruction-followable, and real-time capable model. Its comprehensive coverage, scalable architecture, and broad task alignment make it a significant step toward general-purpose audio intelligence.

\section{Challenges and Future Trends}

Although Kimi-Audio has achieved significant advancements in building universal audio foundation models, several challenges remain in the quest for more capable and intelligent audio processing systems. We describe the challenges and point out several exciting future directions as follows.

\begin{itemize}

\item \textbf{From Audio Transcription to Audio Description.}
Current pre-training paradigms for audio foundation models typically leverage audio-text pre-training to bridge the gap between text and audio, where the text is obtained from audio (speech) by ASR transcription. However, text transcription focuses on the content of spoken words (what is said), neglecting important information in audio, such as paralanguage information (e.g., emotion, style, timbre, tone), acoustic scene, and non-linguistic sounds. Thus, it is important to introduce descriptive text (i.e., audio caption) to depict the audio in richer context. Incorporating both transcriptive and descriptive text of the audio enables models to better understand and generate not only spoken language but also complex acoustic environments, paving the way for more nuanced, multimodal audio processing systems, and thus more general and versatile audio intelligence.

\item \textbf{Better Audio Representations.}
Current audio leverages semantic tokens or acoustic tokens as its representations. Semantic tokens are typically obtained by ASR-based auxiliary loss, which focuses on transcription-oriented information and fails to capture rich acoustic details crucial for understanding and generation. Acoustic tokens are typically learned by audio reconstruction loss, which focuses on description-oriented acoustic details and fails to capture abstractive semantic information that is crucial to bridge to text intelligence. A valuable research direction is to develop representations that integrate both transcription-oriented semantic information and description-oriented acoustic features, encompassing nuances like speaker identity, emotion, and environmental sounds while maintaining high-level abstractive information, which is paramount for more sophisticated audio understanding and generation.

\item \textbf{Throw Away ASR and TTS in Audio Modeling.}
Current audio foundation models rely heavily on ASR and TTS to generate training data in both the pre-training and fine-tuning stages. The quality of training data is constrained by the text recognition accuracy of ASR and expressiveness/diversity/quality of synthesized speech in TTS. In this way, the audio models behave like a sophisticated distillation of existing ASR and TTS systems. As a result, they can hardly achieve performance far beyond the ceiling of ASR/TTS and cannot achieve truly autonomous audio intelligence. An important future direction is to train audio models without relying on ASR/TTS-based pseudo audio data but relying on native audio data, which can result in much higher performance upperbound.

\end{itemize}

\printbibliography[title={References}]

\newpage
\appendix

\section*{Appendix}

\section{Contributions}

\begin{multicols}{2}
\noindent

\textbf{Core Contributors} \\

Ding Ding\\
Zeqian Ju\\
Yichong Leng\\
Songxiang Liu\\
Tong Liu\\
Zeyu Shang\\
Kai Shen\\
Wei Song\\
Xu Tan${^\#}$\\
Heyi Tang\\
Zhengtao Wang\\
Chu Wei\\
Yifei Xin\\
Xinran Xu\\
Jianwei Yu\\
Yutao Zhang\\
Xinyu Zhou${^\#}$\\

\vspace{5em}
\textbf{Contributors} \\

Y. Charles\\
Jun Chen\\
Yanru Chen\\
Yulun Du\\
Weiran He\\
Zhenxing Hu\\
Guokun Lai\\
Qingcheng Li\\
Yangyang Liu\\
Weidong Sun\\
Jianzhou Wang\\
Yuzhi Wang\\
Yuefeng Wu\\
Yuxin Wu\\
Dongchao Yang\\
Hao Yang\\
Ying Yang\\
Zhilin Yang\\
Aoxiong Yin\\
Ruibin Yuan\\
Yutong Zhang\\
Zaida Zhou\\
\end{multicols}

${^\#}$ Project leads. \\
The contributor list is in alphabetical order based on their last names.

\end{document}